\documentclass[letterpaper, 10 pt, conference]{ieeeconf}  % Comment this line out if you need a4paper

\IEEEoverridecommandlockouts                              % This command is only needed if 
\usepackage{algorithmicx}
\usepackage{algpseudocode}
\usepackage{pseudocode}
\usepackage{algorithm}
\usepackage{epsfig} % for postscript graphics files
\usepackage{graphicx} % for pdf, bitmapped graphics files
\usepackage{amsmath} % assumes amsmath package installed
\usepackage{amssymb}  % assumes amsmath package installed
\usepackage{mathtools}
\usepackage{enumerate}
\usepackage{subfig}
\usepackage{float}
\usepackage{cite} % get sequential citations as ranges
\usepackage{balance}
\usepackage{stfloats}
\usepackage{epstopdf}
\usepackage{multirow}
\usepackage{tabularx}

\title{\LARGE \bf
Can a Robot Hear the Shape and Dimensions of a Room?
}

\author{Linh Nguyen$^{1}$, Jaime Valls Miro$^{1}$ and Xiaojun Qiu$^{2}$% <-this % stops a space
\thanks{$^{1}$Linh Nguyen and Jaime Valls Miro are with Centre for Autonomous Systems, University of Technology Sydney, Ultimo, New South Wales 2007, Australia
        {\tt\small vanlinh.nguyen@uts.edu.au} and {\tt\small jaime.vallsmiro@uts.edu.au}}%
\thanks{$^{2}$Xiaojun Qiu is with Centre for Audio, Acoustics and Vibration, University of Technology Sydney, Ultimo, New South Wales 2007, Australia {\tt\small xiaojun.qiu@uts.edu.au}}}%

\begin{document}

\maketitle
\thispagestyle{empty}
\pagestyle{empty}

%%%%%%%%%%%%%%%%%%%%%%%%%%%%%%%%%%%%%%%%%%%%%%%%%%%%%%%%%%%%%%%%%%%%%%%%%%%%%%%%
\begin{abstract}
Knowing the geometry of a space is desirable for many applications, e.g. sound source localization, sound field reproduction or auralization. In circumstances where only acoustic signals can be obtained, estimating the geometry of a room is a challenging propostition. Existing methods have been proposed to reconstruct a room from the room impulse responses (RIRs). However, the sound source and microphones must be deployed in a feasible region of the room for it to work, which is impractical when the room is unknown. This work propose to employ a robot equipped with a sound source and four acoustic sensors, to follow a proposed path planning strategy to moves around the room to collect first image sources for room geometry estimation. The strategy can effectively drives the robot from a random initial location through the room so that the room geometry is guaranteed to be revelaed. Effectiveness of the proposed approach is extensively validated in a synthetic environment, where the results obtained are highly promising.
\end{abstract}

%%%%%%%%%%%%%%%%%%%%%%%%%%%%%%%%%%%%%%%%%%%%%%%%%%%%%%%%%%%%%%%%%%%%%%%%%%%%%%%%
\section{INTRODUCTION}
Room geometry including shape and wall locations or dimension of a room plays a very crucial role in many applications. For instance, geometrical characteristics of a room can ameliorate accuracy of results for applications such as indoor sound source localization \cite{Nguyen2019a,Nguyen2019b}, sound field reproduction \cite{Betlehem2005} and mapping a 3D sound source in autonomous robotic systems \cite{Su2016}. Likewise, in other applications including teleconferencing, virtual reality and auralization, geometrical information of a room can be exploited to create a hallucination or counterbalance effect of room reverberations \cite{Dokmanic2013}. 
%Nevertheless, in many of those applications, where only acoustic signals can be obtained, the room geometry is required to be directly estimated from the sound information.
Particularly for scenarios where no other sensor modalities (e.g. laser range finders or cameras) can be used, infering room geometries directly from acoustic signals is desirable.

\begin{figure}[tb]
\centering
	\includegraphics[scale=0.45]{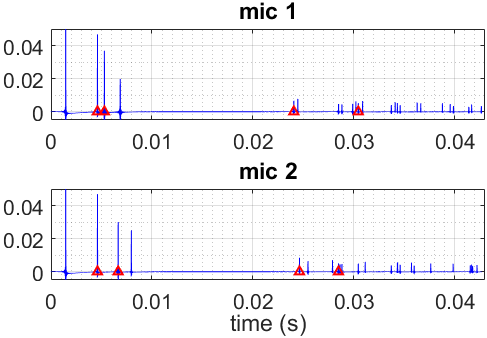}
	\caption{Room Impulse Response (RIRs), where Time of Arrivals (TOAs) from any first Image Source (IS) to microphone~\textbf{i} depicted in red.}
	\label{motivation}
	\vspace{-4mm}
\end{figure}

In recent years, several methods have been proposed to estimate the shape of a room from the room impulse response (RIR) \cite{Jager2016, Coutino2017, Crocco2017, Crocco2018}. For instance, Dokmani\'c~\textit{et al.}~\cite{Dokmanic2013} employed properties of Euclidean distance matrices to localize image sources (ISs)~\cite{Allen1979} of the real source (virtual sound sources located behind a reflective wall at a distance equal to the distance of the source from the wall), which are then used to estimate room shape. Jager~\textit{et al.}~\cite{Jager2016} proposed to formulate room reconstruction as a maximum independent set problem, where combinations of echoes are presented at nodes in a graph, and solved by graph theory. The the computational complexity of their work was improved by Coutino \textit{et al.}~\cite{Coutino2017}, who exploited a subspace based filtering and proposed the use of a single source and a greedy algorithm to find feasible combinations of reflections. Likewise, the work of Crocco~\textit{et al.}~\cite{Crocco2018} used a greedy algorithm to address the NP-hard echo labelling problem for room shape estimation in polynomial time. In previous work by some of the authors~\cite{Rajapaksha2016}, an intuitive geometrical approach was proposed to identify the first ISs before estimating the room geometry.

In addition to static deployment of source and microphones in the aforementioned works, a mobile robot has been exploited in several attempts to reconstruct room shapes. In~\cite{Peng2015,Wang2016a}, the authors utilized a collocated loudspeaker and microphone mobile node to collect first order echoes, which are then employed to reconstruct room shape. Similarly, Krekovi\'c~\textit{et al.}~\cite{Krekovic2016} proposed to use a robot to record room impulse responses (RIRs) when it travels on a predefined path in a room. They then employed the fast simultaneous localization and mapping algorithm to establish the room shape. It is telling, however, that none of these works has considered how the acoustic mobile device should navigate around the room. Moreover, as remarked in~\cite{Dokmanic2013,Wang2016a}
%, in order to successfully recover the shape of a room through sound analysis, 
the sound source and acoustic sensors must be located  in a feasible region - a point where all the first echoes of the emitted signal can be received. However, since the room is initially unknown, finding these regions~\textit{a priori} is impractical. 
%Moreover, due to noise or the acoustic source and microphones possibly located in an infeasible region, 
Should this not be observed (e.g. in the corners of a room), it is often the case that higher order ISs come before first ISs in the RIR signals, as shown in Fig. \ref{motivation}, an effect that is only but compounded by signal noise. 

Addresing the path planning problem for a robot carrying an acoustic package able to emit and record sound signals is thus proposed in this work to recover the shape of a room. Taking advantage of the robot movements, the proposed scheme does not rely on having to collect all the first ISs corresponding to all the walls  in a room at each single step. This, in turn, allows for the the sound source and acoustic sensor device to be deployed at any location in the room. A navigational algorithm is proposed  to autonomously direct the robot around the room in a manner that can most effectively accumulate geometrical characteristics of the room  over time, so as to establish the first ISs of an acoustic source and therefore guarante successful recovery of the room shape.
% LINH: ``guarantee:'' is a strong word, pls make sure ths is indeed made explicit in the formulation of the planner. If the 3 stop will always guarantee we find the first IS, uniquely, then this should be ok.  

The paper is arranged as follows. Section~\ref{sec_2} delineates the problem of acoustically estimating room geometries. The suggested sampling algorithm producing robot paths is discussed in Section~\ref{sec_3}. Simulation results are presented in Section~\ref{sec_4} before conclusions are drawn in Section~\ref{sec_5}.

\section{PROBLEM STATEMENT}
\label{sec_2}
%\subsection{Assumptions}
Omnidirectionality of the sound source is assumed in this work, as well as the existance of an array $K$ of omnidirectional acoustic microphones. %sensors.% When the source emits sounds to surrounding environment, the sensors record sound signals, which consist of characteristics of the room geometry. 
%By processing the microphone recordings, the room geometry can be estimated. 
More sensor unit details will be provided in Section~\ref{Sec_sensor}.
The discussion will focus on 2D scenarios, leaving the extension to 3D for future work. %, perceived to be reasonably straightforward,
Moreover, the actual extraction of RIRs from the sound recordings is outside the scope of this work. The interested reader is referred to~\cite{Farina2000} for more details of a well-known technique.
% for measuring RIRs. 
Finally, it can be safely assumed that the microphones can be kept from colliding with the room walls while the robot is running.

\subsection{Modelling}
It has been established in the literature that the shape of a room can be acoustically obtained from the corresponding room impulse responses (RIRs), which can be extracted from sound signals recorded by microphones~\cite{Dokmanic2013, Rajapaksha2016, Krekovic2016}. Theoretically, RIRs contain direct sound from the source, arbitrary reflections of sound or echoes and measurement noises. In other words, sound propagation can be seen as a family of RIRs, and at each acoustic sensor RIR can be mathematically represented by
\begin{equation}
h_k(t)=\sum_{i\geq 0}\beta_i\delta(t-\tau_{k,i})+\epsilon(t), \;\; k\in\{1,\cdots, K\}
\label{equ_1}
\end{equation}
where $\epsilon(t)$ is measurement noise, which is independent from the signals at each acoustic sensor. $\beta_i$ is a factor that is dependent of the absorption coefficient of the wall $i>0$. $\delta$ is the Dirac delta impulse induced by the real sound source or the reflections/echoes. $\tau_{k,i}$ is the propagation time or the time of arrival (TOA) from the sound sources or the reflections/echoes to the microphone $k$, specified by
\begin{equation}
\tau_{k,i}=\frac{\parallel I_{*,i}-M_k\parallel}{c},
\label{equ_2}
\end{equation}
where $I_{*,i}$ is the location of any image source as a mirror of the real source across the wall $i$. $M_k$ is the location of the microphone $k$ while $c$ and $\parallel\cdot\parallel$ denote the speed of sound propagation and the Euclidean distance, respectively. It is noticed that in (\ref{equ_1}), when $i=0$, the elements of RIRs directly come from the acoustic source.

\begin{figure}[tb]
\centering
	\includegraphics[scale=0.349]{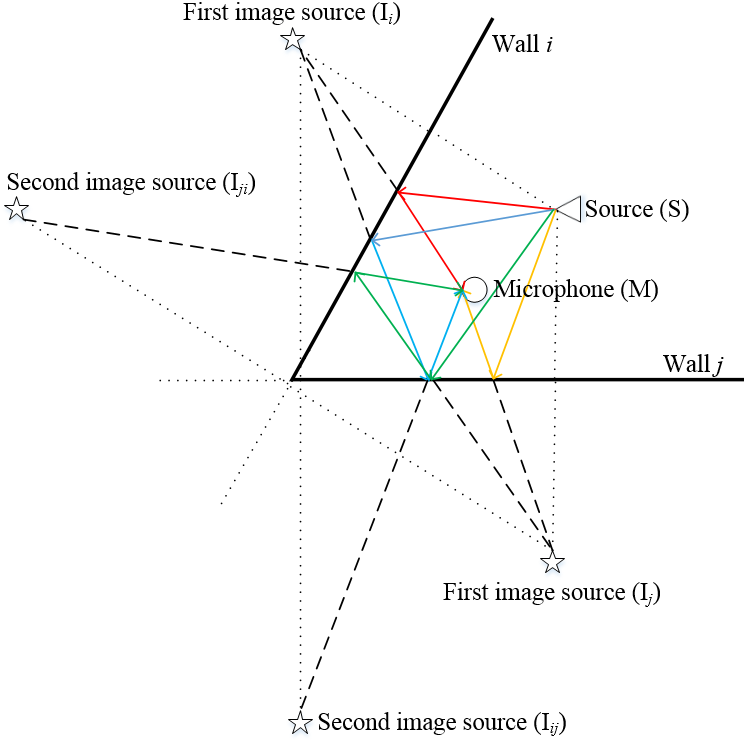}
	\caption{Principle of image sources.}
	\label{IS_principle}
	\vspace{-4mm}
\end{figure}

Moreover, the reflection of sound, or echo, can be modelled by an image source (IS)~\cite{Allen1979}, which is considered as a mirror image of the real acoustic source across a corresponding wall, as illustrated in Fig. \ref{IS_principle}. It can be seen how there are first and higher order image sources (ISs). In an idealistic scenario, the first IS ($\text{I}_i$) is a mirror image of the source (S) across the wall $i$. Likewise, the second IS ($\text{I}_{ij}$) is a mirror image of the first IS ($\text{I}_i$) across the wall $j$. Physically, the first IS ($\text{I}_i$) can be explained as a reflection of sound from the source (S) on the wall $i$ to the microphone (M) while the second IS ($\text{I}_{ij}$) can be accounted for the reflection of sound from the source (S) on the wall $i$ and then on the wall $j$ before reaching to the microphone (M), as shown in the red lines and light blue lines in Fig. \ref{IS_principle}, respectively.

Therefore, we can now consider the Time Of Arrivals (TOA) as the time it takes for the sound to travel from either the real source or ISs to a microphone. In other words, if TOA is known, a distance from the microphone to either the real source or corresponding IS can be inversely computed, given $c$. It is apparent that if the RIRs are measured, the TOAs can be easily obtained by simply processing the impulse signals, e.g. picking the signal peaks over time. On the other hands, there is a link between ISs and the room geometry: if the location of the first IS is known, given the real source location, a corresponding wall line can be found. However, knowing TOAs and the distances from ISs to the microphones does not guarantee knowing the locations of ISs since the distances from ISs to the microphones are unlabelled. Moreover, due to noise, acoustic sensors record reflections/echoes in an arbitrary order. For instance, as shown in Fig. \ref{motivation}, the fourth peak of all the RIRs, which does not correspond to any first IS, arrives earlier than the two other first ISs. Similarly, for mic 1, the fourth of the first ISs arrives at the eleventh peak, while on mic. 2 arrives at the eighth peak of their RIRs, yet they represent the same wall (line) in the room. How to find the first IS from the RIRs is thus critical, and is discussed next.

\subsection{Common Image Sources}
\label{2_b}
Given the TOA from RIRs and then a distance $d$ from a possible IS to a microphone, there are infinite possible locations of the IS at that distance: the locations on a circle centred at the microphone location with a radius of distance $d$. Furthermore, if there is only one microphone, it is not possible to find the first IS corresponding to a wall. For instance, on the RIR of mic. 1 in Fig.~\ref{motivation}, the first ISs arrive arbitrarily, corresponding to the second, third, fifth and eleventh peaks, respectively. Therefore, in practice, it has been proposed to employ multiple acoustic sensors for room shape estimation~\cite{Dokmanic2013}, and in this work four microphones are employed to reconstruct the shape of a 2D polygonal room~\cite{Rajapaksha2016}.

Let us consider two RIR signals as demonstrated in Fig. \ref{motivation}. For each RIR, there are multiple potential but unlocated corresponding ISs. If one can smartly combine those ISs, it is possible to find a location of a common IS from all the RIR signals, which would corresponds to a wall in the room. If we define $I=(x,y)$ as a location of any IS and $M_k=(x_k,y_k)$ as the location of the microphone $k$, $k\in\{1,\cdots,4\}$, then a location for the common IS can be derived from (\ref{equ_2}) by solving
\begin{equation}
I\times\textbf{M}=\textbf{b},
\label{equ_3}
\end{equation}
where
\begin{equation}
\textbf{M}=
\begin{bmatrix}
(M_1-M_2)^T	\\[0.3cm]
(M_2-M_3)^T \nonumber
\end{bmatrix},
\end{equation}
\begin{equation}
\textbf{b}=0.5\begin{bmatrix}
x_1^2-x_2^2+y_1^2-y_2^2-(\tau_1c)^2+(\tau_2c)^2	\\[0.3cm]
x_2^2-x_3^2+y_2^2-y_3^2-(\tau_2c)^2+(\tau_3c)^2 \nonumber
\end{bmatrix}.
\end{equation}
Here $\tau_1$, $\tau_2$ and $\tau_3$ are one of the TOAs on the RIRs collected by mc.i 1, 2 and 3, respectively. As discussed in our previous work~\cite{Rajapaksha2016}, an additional fourth acoustic sensor is employed to verify the common ISs. In other words, two other equations similar to (\ref{equ_3}) can be established and solved, then three sets of solutions are exploited to sort out the common ISs. In practice, any IS location close to its counterpart, which is less than a pre-defined small distance, is kept for further processing.

Moreover, the common IS can be further verified to be a realistic IS, or derive from noise, by the use of a reflective point, i.e. the intersection between the line from the IS to a microphone and a potential wall line. If the reflective point and the real sound source are in different sides of a reconstructed or potential wall line, the corresponding IS is treated as noise and removed from the list of common ISs.

\begin{figure}[tb]
\centering
	\includegraphics[scale=0.45]{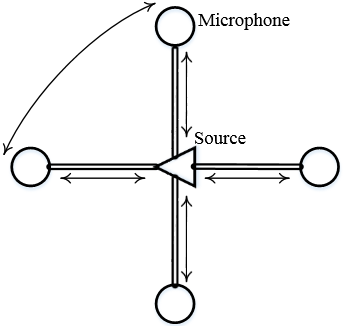}
	\caption{Robot mechanism for collecting sound signals.}
	\label{Sensing_config}
	\vspace{-4mm}
\end{figure}

\subsection{Room Shape Reconstruction}
Theoretically, if an acoustic source and accompanied microphones are located in a feasible region of a room~\cite{Dokmanic2013, Wang2016a}, the several common ISs that are located closer to the real sound source than any other ISs are regarded as the fist ISs.% present to all the potential wall lines of the room. 
% Linh - `which present to all' ??? I dont understand this sentence -  
Once locations of the first ISs are computed, the corresponding wall lines can be easily established as illustrated in Fig.~\ref{IS_principle}, and the room can then be reconstructed according to the following theorem:

\textbf{Theorem 1:} A 2D polygonal room can be constructed if and only if any an established wall line has two intersections with other established wall lines.

\begin{proof}
Geometrically, the proof is straightforward.
\end{proof}

Therefore, this work aims to find the first ISs of the real acoustic source to reconstruct the room geometry. A path planning strategy is next proposed whereby common ISs are accumulatively collected and verified over time.

\section{ROBOT MOTION PLANNING} %SAMPLING STRATEGY FOR A ROBOT TO CONSTRUCT A ROOM}
\label{sec_3}
%To effectively localize the first ISs of the source a path planning strategy is proposed whereby common ISs are accumulatively collected and verified over time. We will discuss the proposition in details in this section.

\subsection{Sensing Configuration}
\label{Sec_sensor}
The robotic sensing unit proposed in this work to emit and record the sound signals is demonstrated in Fig.~\ref{Sensing_config}. 
%As discussed in Section~\ref{sec_2}, to infer the room geometry in 2D scenarios, it is proposed to employ one source and four acoustic sensors.
In this design, the acoustic source is embedded on a robot while the four microphones are installed on four robot arms, which can rotate around the robot, hence allowing the microphones to be located at different positions~\cite{Rajapaksha2016}. Since distance between source and microphone plays a considerable role in forming the spectrum of RIRs, it is also proposed that the microphones can be flexibly extend along the robot arms, hence being able to adjust its distance to the source as required. This design enables the system to collect more information on geometrical characteristics of the room even when the robot is stationary. 
This reconfigurabiliy 
%Flexibility of the designed mechanism would benefit the system, where it is consistently able to find a highly potential 
increases the chances of finding first IS as the robot moves to any location in the room. This is particularly effective to overcome issues when the robot moves to a corner, as discussed in more detail in Section~\ref{Sec_corner}.

%It is noted that the mechanism in Fig. \ref{Sensing_config} is designed to effectively collect sound signals in a room, yet the actual extraction of RIRs from the sound recordings is outside the scope of this work. The interested reader is referred to~\cite{Farina2000} for more details of a well-known technique for measuring RIRs. Moreovery, it can be safely assumed that the microphones can be kept from colliding with the room walls while the robot is running.

\subsection{Clustering}
\begin{figure}[t]
\centering
	\includegraphics[scale=0.28]{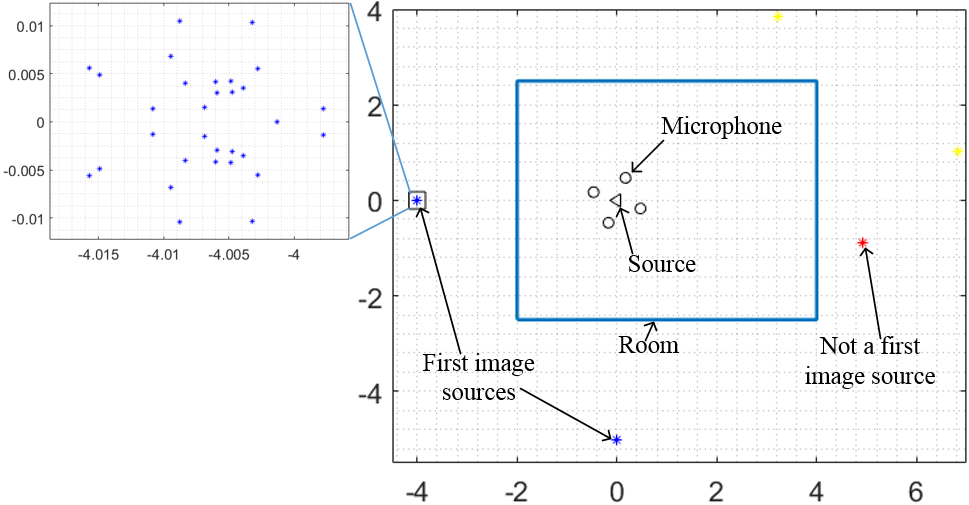}
	\caption{Robot arm rotation comparisons: without rotation, the system reports the wrong common first IS (red and yellow stars); 
	with rotation, the system reports a set of 31 approximate common first ISs (blue stars).}% Axis units are in meters.}
	\label{Clustering}
	\vspace{-4mm}
\end{figure}

By taking advantage of the rotating arms configuration %in the proposed mechanism as shown in Fig.~\ref{Sensing_config}, 
the system is equipped with more opportunities to guarantee the common first ISs. In other words, given a fixed sound source, at every set of the microphone locations, the system reports a set of the potential common first ISs. When the microphones rotate through multiple orientations, the multiple sets of the potential common first ISs are grouped and then clustered by using a nearest neighbour technique. The biggest cluster, which is closest to the real sound source as compared to other possible clusters, is considered as the group of correct common first ISs obtained by the acoustic sensors.
This can be seen in the example demonstrated in Fig.~\ref{Clustering}. %When the system of a sound source and four acoustic sensors was deployed as shown in Fig. \ref{Clustering},
The common ISs reported are located at points marked by the red and yellow stars. 
% Linh - yellow stars are hardly visible!
It can be clearly seen that those common ISs are not the expected first ISs. Nonetheless, when the microphones were rotated around the acoustic source (every 10 degrees), at every stop, the system reports the common ISs. After 36 iterations, all the common ISs are clustered, and the biggest group including 31 common ISs (blue stars, approximately located at the point (-4, 0) ), would be the nominated location of the first would-be IS.%, and would be nominated for the constructing a wall line of a room.

\begin{figure}[t]
\centering
	\includegraphics[scale=0.35]{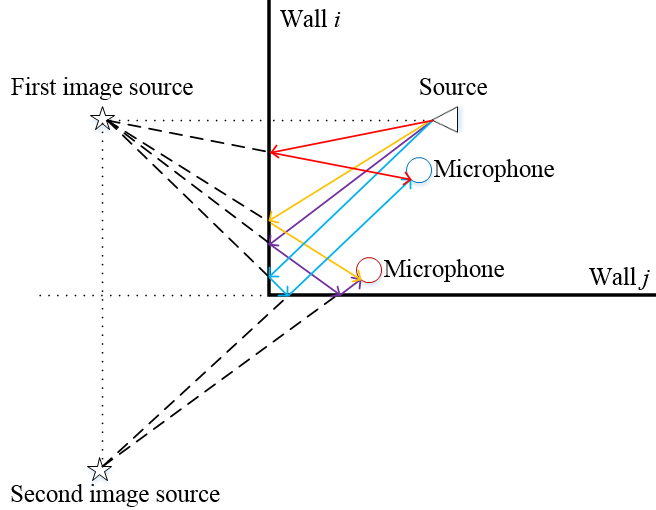}
	\caption{Propagation of sound in room corners.}
	\label{Corner_sound_pro}
	\vspace{-4mm}
\end{figure}

\subsection{Corner Issues}
\label{Sec_corner}
\begin{figure*}[tb]
\centering
	\subfloat[]{\label{corner_bad}
		\includegraphics[scale=0.3]{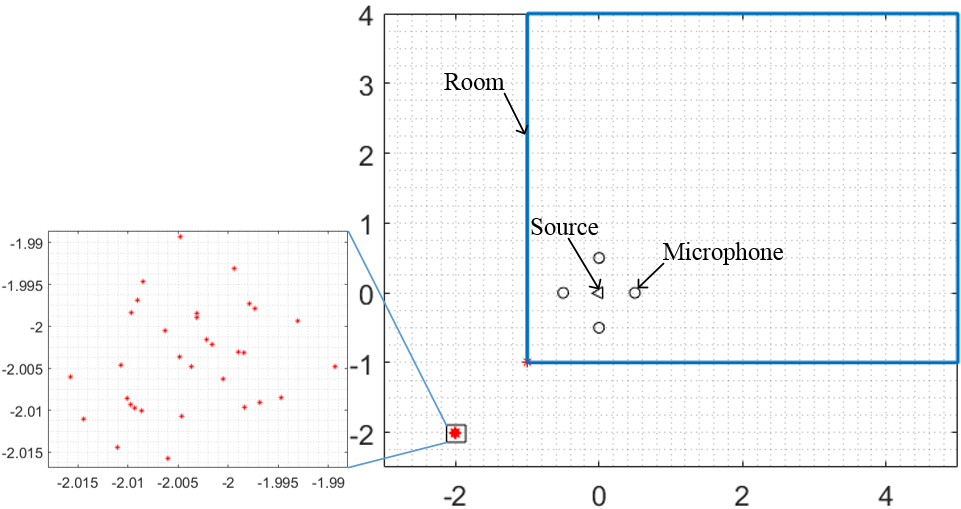}} \hspace{3em}
	\subfloat[]{\label{corner_good}
		\includegraphics[scale=0.3]{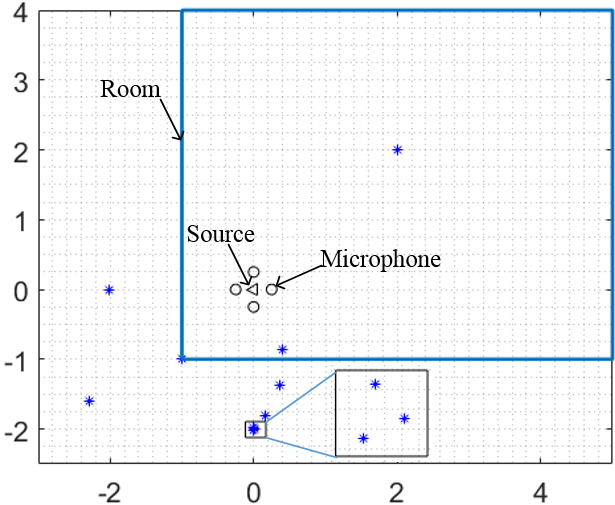}}
	\caption{Issues around corners: (a) the microphones are far from the source and cannot find common first ISs and (b) the microphones move closer to 
	the source and be able to find common first ISs. Axis units are in meter.}
	\label{Corner_issue}
	\vspace{-4mm}
\end{figure*}

Since the room geometry is unknown, it is likely that a mobile robot may travel through a corner of the room. In this case, when a microphone is located close to the walls a issue appears where the TOAs from the first or higher order ISs to the microphone are very similar. An illustrative example is shown in Fig.~\ref{Corner_sound_pro}, where the TOA from the source to the red microphone reflecting exclusively on wall $i$ (corresponding to the first IS, depicted in orange), and that from the source reflecting on both walls $i$ and $j$ (corresponding to the second IS, depicted in purple), are very close. Under the influence of noises, the system may mistakenly report the second IS rather than the expected first IS as the common ones. A simulation example is illustrated by Fig.~\ref{corner_bad},
%ticularly, if both of the source and microphones were located as
whereby by only rotating the acoustic sensors around the source and clustering the common ISs, the system reported the best location for the most common ISs (red stars) around  (-2, -2), which is not a location of any first IS for the real source.
To address the issue, it is proposed to simultaneously move the microphones along the robot arms to different locations, hence changing the distance between source and the acoustic sensors and resulting in different RIRs. When the microphones become closer to the acoustic source, as seen in Fig.~\ref{corner_good}, the system was able to reported many common ISs (blue stars), with the most common ISs approximately located at (0, -2), the actual location of the fist IS of the acoustic source.

\subsection{Sampling Path}
This section will discuss a sampling strategy for the robot to navigate in unknown environments so that the room geometry can be recoverd in an effective manner. While a robot has been used in the literature for the purpose of room shape estimation~\cite{Krekovic2016, Wang2016a} while traversing on a predefined path, planning in new environmets that has not been investigated yet. It is noted that since the room is unknown, it is presumed that origin of the coordinate is always at the initial location of the robot.

The premise behind the strategy is that to establish new line wall in a room is to drive the robot to at least three stops. At each stop the system %including a sound source and four microphones 
would report a location of the common first ISs as described thus far in the paper, representing the largest clusters and closest to the real source as compared with the other possible clusters in that iteration. Theoretically, if a first IS is found, given two locations of both the source and its first IS, a wall line of a room, which is perpendicular to the line from the source to the first IS, is uniquely formed. However, given the noisy RIRs as captured by the acoustic sensors, and the erroneousness in processing the sound signals, the system may indeed report the wrong common first IS. Hence, the common ISs coming from a single stop are not certain to correspond with a real wall in the room. Furthermore, extensive experimentation has also revealed that in some scenarios, especially when the robot travels to ``infeasible" regions such as corners of the room, and even with  two stops of the robot, the system can still report the common second ISs, matching to the wrong room wall. Therefore, a more robust three-stop strategy for the robot is suggested, where the common ISs are employed to verify the established wall line obtained from the two previous robot stops.

\begin{figure*}[tb]
\centering
	\subfloat[]{\label{pa}
		\includegraphics[scale=0.29]{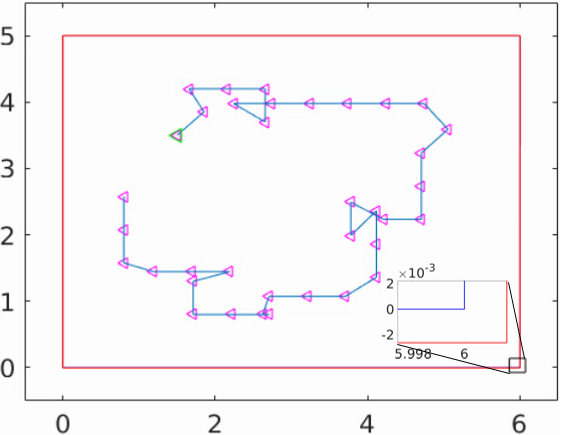}} %\hspace*{0.6em}
	\subfloat[]{\label{pb}  
		\includegraphics[scale=0.29]{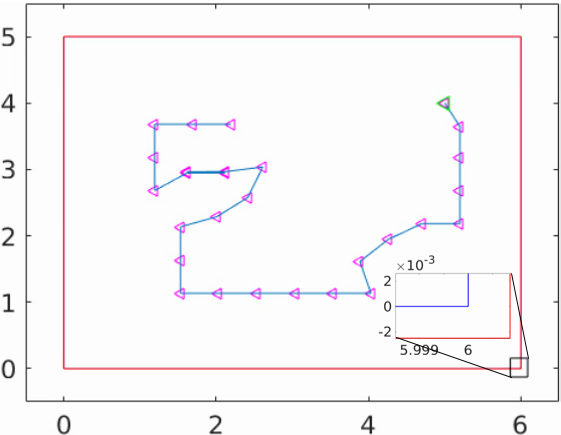}} 
	\subfloat[]{\label{pc}
		\includegraphics[scale=0.29]{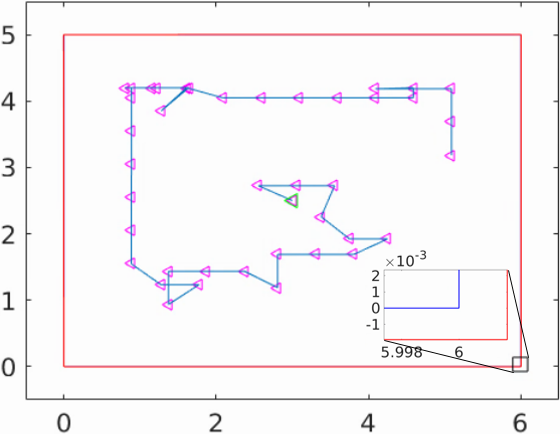}} %\hspace*{0.6em}
	\subfloat[]{\label{pd} 
		\includegraphics[scale=0.29]{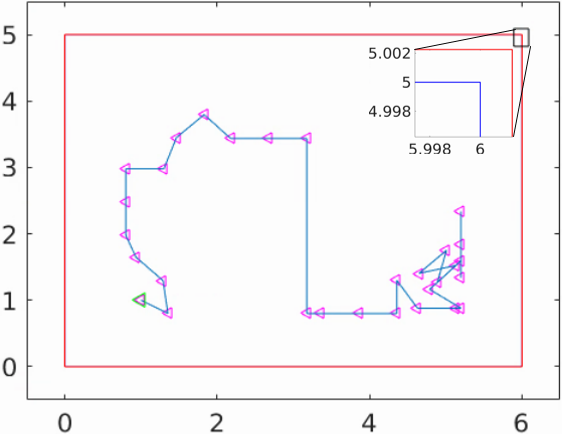}} 
	\caption{Some illustrations of the robot movements to collect acoustic signals, successfully estimate the room shape and compute the room dimension. 
	The initial robot location is marked by a red triangle. Axis units are in meter.}
	\label{Sampling_path}
	\vspace{-4mm}
\end{figure*}

The process to drive the robot to three stops to form a wall line is as follows: at the first robot stop, it is assumed that a location of the common first ISs is found, and a wall line $w_1$ is obtained. It is then proposed that the robot should move on a line that is parallel to the obtained wall line $w_1$ for a given distance. However, there are always two possible directions the robot should move on that parallel line, hence two possible locations for the next stop of the robot. In the proposed algorithm it is proposed that the robot will move to the possible location that is further to the already established or examined walls in the room, effectively prefering room exploration over exploitation in the search. Moving the robot along a line parallel to the obtained wall line $w_1$ aims to increase the chance to find the common first ISs for the same wall line $w_1$. At the second robot stop, if another wall line $w_2$ is obtained from its corresponding common first ISs and approximate to $w_1$, the robot should move to a third stop to verify them. The robot will thus be instructed to drive along a line parallel to $w_2$ to a similar predefined distance. % LINH ???:  but in one direction moving forward from first then second stops. 
At the third robot stop, if the system is able to form a wall line $w_3$ that is approximate to both $w_1$ and $w_2$, then a real wall line of the room is established and confirmed.

In the cases where, at the second or third robot stops, the system cannot find a wall line that is approximate to $w_1$ or both $w_1$ and $w_2$, the distance between the source and the microphones can be modified % to a given value by moving the sensors along the robot arms. 
At each change, the system reports a new wall line and compares it with $w_1$ or both $w_1$ and $w_2$. If approximate matches can be found, %as aforementioned, 
the system proceeds as normal. Otherwise, the robot moves randomly to a new location and the system restarts the searching procedure.

\begin{figure}[t]
\centering
	\subfloat[]{\label{e_f}
		\includegraphics[scale=0.31]{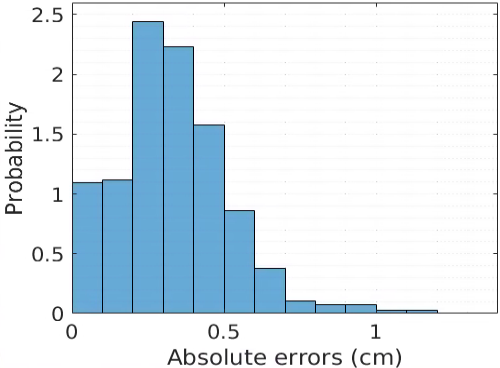}} %\hspace*{0.6em}
	\subfloat[]{\label{e_r} 
		\includegraphics[scale=0.31]{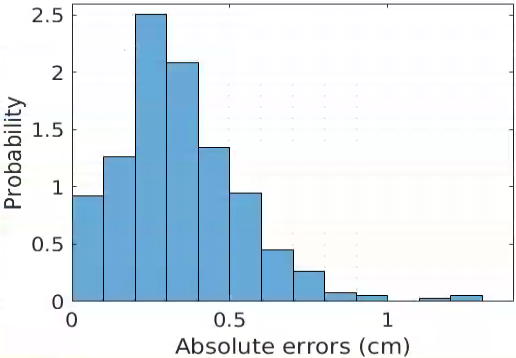}}
	\caption{Histograms of the absolute errors in the room dimension estimation: (a) in the fixed room scenario and (b) in the random room scenario.}
	\label{error}
	\vspace{-4mm}
\end{figure}

\begin{figure}[t]
\centering
	\subfloat[]{\label{s_f}
		\includegraphics[scale=0.3]{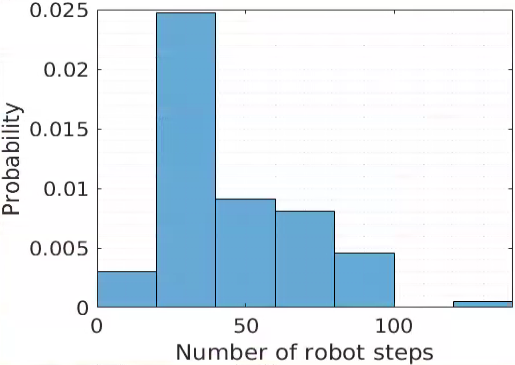}} %\hspace*{0.6em}
	\subfloat[]{\label{s_r}
		\includegraphics[scale=0.3]{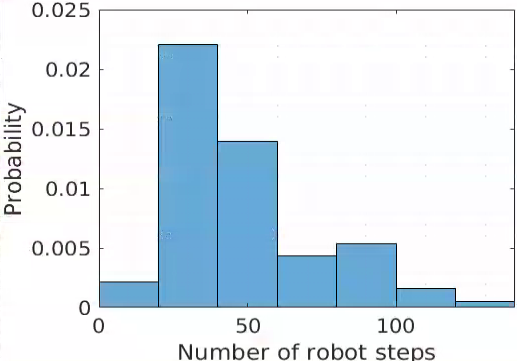}}
	\caption{Histograms of the robot steps required to construct the room geometry: (a) in the fixed room scenario and (b) in the random room scenario.}
	\label{steps}
	\vspace{-4mm}
\end{figure}

It is important to understand the definition of wall approximation for $w_1$, $w_2$ and $w_3$. Two wall lines are approximate when their angles to the horizontal line remain  similar. To avoid being parallel, two perpendicular vectors projected from the initial robot location to them should have the same directions and approximate magnitudes.

It should also be noted that after a wall line is found and confirmed, the proposed approach utilizes the second biggest cluster of the common ISs in the current robot stop to establish the next wall line for the room, following the aforesaid procedure. The room is fully constructed when any an established wall line has two intersections with other established wall lines as proven by \textbf{Theorem 1}.

To examine the proposed algorithm please refer to the examples in Fig.~\ref{Sampling_path}. There are three scenarios where the robot starts from a random locations as shown in Figures \ref{pa}, \ref{pb} and \ref{pd}, while in Fig. \ref{pc} the robot start from the middle of the room. In all these four illustrations, the system was able to effectively find the room shape, localize the wall lines and compute the dimension. The real wall lines of the room are plotted in blue, while the estimates are framed in red lines. The estimation error is smaller than 1 cm. It can also be noticed that in these figures, though the origin of the original coordinate was located at the initial robot location as delineated in the proposed algorithm, all plots were transformed to a common coordinate for a consistent presentation.

\section{SIMULATION RESULTS AND DISCUSSION}
\label{sec_4}
To demonstrate effective of the proposed approach, we extensively carried out different experiments on a synthetic simulation environment by using an RIR generator~\cite{Habets2010online}, and an implementation of the acoustic image method~\cite{Allen1979}. In the simulated room, the reverberation time was set to 0.8 s while the sampling rate and the sound propagation speed were set to 96 kHz and 343 m/s, respectively. The sensing configuration shown in Fig.~\ref{Sensing_config} was employed. The extension distance between the real sound source and the acoustic sensors along the robot arms could reach to a maximum of 0.5 m. The maximum distance the robot could move at each step was set to 0.5 m.

The RIRs were first recorded by the microphones and then processed to pick their peaks over time, which were employed to compute the TOAs. Each TOA corresponds to a potential IS and the common ISs can be found by following the procedure described in Section~\ref{2_b}.

During the experiments the room geometry was unknown. In other words, the initial location of the robot was random; and due to this randomness, we repeated the experiments at each simulation 1000 Monte-Carlo trials. Two scenarios were considered: (i) the simulated room was sized by 6 m $\times$ 5 m, and (ii) the simulated room was randomly sized.

After running the simulations, in all the experiments, the robot system was able to find the correct rectangular room shape. That is, at each experiment the robot stopped running when four wall lines were found. More importantly, the proposed algorithm successfully localized the wall lines in the room, resulting in highly accurate room dimensions as  illustrated by Fig.~\ref{error}, depicting the absolute errors between the real wall lines and the estimated wall lines. Absolute errors in the room dimension estimation in both scenarios considered are 95 percent smaller than 1 cm, and mostly around 0.35 cm. It shows that the estimation errors are not dependent on room dimension or the initial location of the robot.
The numbers of the robot steps required in all the experiments to successfully construct the room geometry are summarized in Fig.~\ref{steps}. Similar to the room dimension estimation errors, the number of robot steps in all the experiments is independent to the geometrical characteristics of the room. That is, in about 95 percent of the experimental cases, the robot was required to run less than 100 steps to securely shape the room, where the numbers are most likely at 44 and 48 steps for the paths followed by the obot in the fixed and random rooms respectvely

%Gumbel
%fixed room but random start
%error
%m= 0.43 cm
%s=0.22 cm 
%
%%robot steps
%m=56.62 steps
%s=27.20 steps
%
%%random room and random start
%error
%m= 0.45cm
%s=0.26 cm
%
%%robot steps
%m=61.65 steps
%s=29.83 steps

\section{CONCLUSIONS}
\label{sec_5}
The paper has discussed a new and efficient approach to construct room geometry from acoustic signals only by the use of a mobile robot, a sound source and four microphones. The robot carries the source and acoustic sensors to travel around a room, where the common first ISs are accumulatively extracted from the collected RIRs. Moreover, the robot navigation is driven by an effective path planning  strategy to guarantee the room geometry to be successfully estimated in a reasonable time. The key advantages of the proposed method are that the robot can start from a random location and is applicable to any-sized room. The results obtained in the extensive simulation experiments have demonstrated the efficacy of the proposed algorithm.

\addtolength{\textheight}{-12cm}   % This command serves to balance the column lengths
                                  % on the last page of the document manually. It shortens
                                  % the textheight of the last page by a suitable amount.
                                  % This command does not take effect until the next page
                                  % so it should come on the page before the last. Make
                                  % sure that you do not shorten the textheight too much.

%%%%%%%%%%%%%%%%%%%%%%%%%%%%%%%%%%%%%%%%%%%%%%%%%%%%%%%%%%%%%%%%%%%%%%%%%%%%%%%%

\balance

\bibliographystyle{IEEEtran}
\bibliography{IEEEabrv,References}

\begin{thebibliography}{10}
\providecommand{\url}[1]{#1}
\csname url@rmstyle\endcsname
\providecommand{\newblock}{\relax}
\providecommand{\bibinfo}[2]{#2}
\providecommand\BIBentrySTDinterwordspacing{\spaceskip=0pt\relax}
\providecommand\BIBentryALTinterwordstretchfactor{4}
\providecommand\BIBentryALTinterwordspacing{\spaceskip=\fontdimen2\font plus
\BIBentryALTinterwordstretchfactor\fontdimen3\font minus
  \fontdimen4\font\relax}
\providecommand\BIBforeignlanguage[2]{{%
\expandafter\ifx\csname l@#1\endcsname\relax
\typeout{** WARNING: IEEEtran.bst: No hyphenation pattern has been}%
\typeout{** loaded for the language `#1'. Using the pattern for}%
\typeout{** the default language instead.}%
\else
\language=\csname l@#1\endcsname
\fi
#2}}

\bibitem{Nguyen2019a}
L.~Nguyen, J.~VallsMiro, and X.~Qiu, ``Multilevel {B-S}plines based learning
  approach for sound source localization,'' \emph{IEEE Sensors Journal}, vol.
  19(10), pp. 3871 -- 3881, 2019.

\bibitem{Nguyen2019b}
L.~Nguyen and J.~VallsMiro, ``Acoustic sensor networks and mobile robotics for
  sound source localization,'' in \emph{Proc. IEEE International Conference on
  Control and Automation}, Edinburgh, United Kingdom, July 2019, pp. 1--6,
  \textit{to appear}.

\bibitem{Betlehem2005}
T.~Betlehem and T.~Abhayapala, ``Theory and design of sound field reproduction
  in reverberant rooms,'' \emph{The Journal of the Acoustical Society of
  America}, vol. 117, pp. 2100--2111, 2005.

\bibitem{Su2016}
D.~Su, K.~Nakamura, K.~Nakadai, and J.~VallsMiro, ``Robust sound source mapping
  using three-layered selective audio rays for mobile robots,'' in \emph{Proc.
  IEEE/RSJ International Conferece on Intelligent Robots and Systems}, Daejeon,
  Korea, October 2016, pp. 2771--2777.

\bibitem{Dokmanic2013}
I.~Dokmani\'c, R.~Parhizkar, A.~Walther, Y.~M. Lu, and M.~Vetterli, ``Acoustic
  echoes reveal room shape,'' \emph{Proceedings of the National Academy of
  Sciences of the United States of America}, vol. 110(30), pp.
  12\,186--12\,191, 2013.

\bibitem{Jager2016}
I.~Jager, R.~Heusdens, and N.~D. Gaubitch, ``Room geometry estimation from
  acoustic echoes using graph-based echo labeling,'' in \emph{Proc. IEEE
  International Conference on Acoustics, Speech, and Signal Processing},
  Shanghai, China, March 2016, pp. 1--5.

\bibitem{Coutino2017}
M.~Coutino, M.~B. Moller, J.~K. Nielsen, and R.~Heusdens, ``Greedy alternative
  for room geometry estimation from acoustic echoes: {A} subspace-based
  method,'' in \emph{Proc. IEEE International Conference on Acoustics, Speech,
  and Signal Processing}, New Orleans, LA, USA, March 2017, pp. 366--370.

\bibitem{Crocco2017}
M.~Crocco, A.~Trucco, and A.~D. Bue, ``Uncalibrated 3d room geometry estimation
  from sound impulse responses,'' \emph{Journal of the Franklin Institute},
  vol. 354, p. 8678–8709, 2017.

\bibitem{Crocco2018}
------, ``Room reflectors estimation from sound by greedy iterative approach,''
  in \emph{Proc. IEEE International Conference on Acoustics, Speech, and Signal
  Processing}, Calgary, Alberta, Canada, April 2018, pp. 6877--6881.

\bibitem{Allen1979}
J.~Allen and D.~Berkley, ``Image method for efficiently simulating small room
  acoustics,'' \emph{Journal of Acoustical Society of Ameria}, vol. 65(4), pp.
  943--950, 1979.

\bibitem{Rajapaksha2016}
T.~Rajapaksha, X.~Qiu, E.~Cheng, and I.~Burnett, ``Geometrical room geometry
  estimation from room impulse responses,'' in \emph{Proc. EEE International
  Conference on Acoustics, Speech, and Signal Processing}, Shanghai, China,
  March 2016, pp. 331--335.

\bibitem{Peng2015}
F.~Peng, T.~Wang, and B.~Chen, ``Room shape reconstruction with a single mobile
  acoustic sensor,'' in \emph{Proc. IEEE IEEE Global Conference on Signal and
  Information Processing}, Orlando, FL, USA, December 2015, pp. 1116--1120.

\bibitem{Wang2016a}
T.~Wang, F.~Peng, and B.~Chen, ``First order echo based room shape recovery
  using a single mobile device,'' in \emph{Proc. EEE International Conference
  on Acoustics, Speech, and Signal Processing}, Shanghai, China, March 2016,
  pp. 21--25.

\bibitem{Krekovic2016}
M.~Krekovi\'c, I.~Dokmani\'c, and M.~Vetterli, ``Echoslam: {S}imultaneous
  localization and mapping with acoustic echoes,'' in \emph{Proc. EEE
  International Conference on Acoustics, Speech, and Signal Processing},
  Shanghai, China, March 2016, pp. 11--15.

\bibitem{Farina2000}
\BIBentryALTinterwordspacing
A.~Farina, ``Simultaneous measurement of impulse response and distortion with a
  swept-sine technique,'' in \emph{Audio Engineering Society Convention 108},
  Februay 2000. [Online]. Available:
  \url{http://www.aes.org/e-lib/browse.cfm?elib=10211}
\BIBentrySTDinterwordspacing

\bibitem{Habets2010online}
\BIBentryALTinterwordspacing
E.~A.~P. Habets, ``Room impulse response generator,'' 2010. [Online].
  Available:
  \url{https://www.audiolabs-erlangen.de/fau/professor/habets/software/rir-generator}
\BIBentrySTDinterwordspacing

\end{thebibliography}

\end{document}